\documentclass[12pt]{article}

\newcommand{\be}{\begin{equation}}
\newcommand{\ee}{\end{equation}}
\newcommand{\bea}{\begin{eqnarray}}
\newcommand{\eea}{\end{eqnarray}}

\renewcommand{\theequation}{\arabic{section}.\arabic{equation}}

\begin{document}
\begin{titlepage}

\vspace{1in}

\begin{center}
\Large
{\bf Cosmic Dynamics of Bose-Einstein Condensates}

\vspace{1in}

\normalsize

\large{James E. Lidsey$^1$}

\normalsize
\vspace{.7in}

{\em Astronomy Unit, School of Mathematical 
Sciences,  \\ 
Queen Mary, University of London, Mile End Road, LONDON, E1 4NS, U.K.}

\end{center}

\vspace{1in}

\baselineskip=24pt
\begin{abstract}
\noindent A dynamical correspondence is established 
between positively curved, isotropic, perfect fluid cosmologies and 
quasi-two-dimensional, harmonically trapped Bose-Einstein 
condensates by mapping the equations of motion for both systems 
onto the one-dimensional Ermakov system. 
Parameters that characterize the physical properties of the 
condensate wavepacket, such as its width, momentum and energy, 
may be identified with the scale factor, 
Hubble expansion parameter and energy 
density of the universe, respectively. Different forms of
cosmic matter correspond to different choices for the time-dependent 
trapping frequency of the condensate. The trapping frequency that mimics
a radiation-dominated universe is determined.
\end{abstract}

PACS NUMBERS: 98.80.Cq, 03.75.Kk

\vspace{.7in}
$^1$Electronic mail: J.E.Lidsey@qmul.ac.uk
 
\end{titlepage}


\setcounter{equation}{0}

\def\theequation{\arabic{equation}}

In recent years, analogies  
between various condensed matter systems and different branches of
gravitational physics have been developed. (For reviews, 
see, e.g., Refs. \cite{volovik,nvv}). For example, 
the propagation 
of acoustic waves in an irrotational, inviscid, barotropic fluid 
is formally equivalent to 
that of a massless scalar field on a curved, Lorentzian spacetime 
\cite{unruh}. 
This implies that quantum field theory in curved space can be modeled 
in terms of the 
quantization of the sound 
wave (phonon field). 
Furthermore, it is possible to model a black hole acoustically
in terms of supersonic fluid flow and, in principle, 
quantum effects associated with black hole event horizons may then be studied 
within the context of condensed matter configurations 
\cite{unruh,visser,otherbh}. A related phenomenon is 
that of cosmological particle production arising 
in a time--dependent gravitational field and the description of 
such a process in a dynamical Bose--Einstein condensate 
has recently been investigated \cite{ch,ff1,bc,blv}. 

Identifying a link in this way between 
gravitational and non--gravitational physical systems 
is of great importance. 
To date, attention has focused primarily on the kinematical properties of 
general relativity, such as the existence of event horizons. 
The purpose of this work is to 
note that a dynamical correspondence may also be established between 
isotropic, four--dimensional cosmological models and 
harmonically trapped, quasi--two--dimensional Bose--Einstein  
condensates. The correspondence arises because the equations 
of motion for both  
systems can be mapped onto the 
one--dimensional Ermakov system \cite{em,pinney}. 
Ermakov systems arise in a variety of physical 
phenomena \cite{ermakovsystem,hl,wk} 
and the one--dimensional case corresponds to 
a second--order, non--linear, ordinary 
differential equation (ODE) \cite{em,pinney}.  

We begin by considering the 
condensed matter system. 
A Bose--Einstein condensate is the ground state of a collection 
of interacting bosons trapped by an external potential. 
In the limit where the number of atoms is sufficiently large 
and the atomic interactions are sufficiently weak, the 
mean--field approximation 
may be invoked, where the effect felt by a particular atom due to the 
ensemble 
is approximated by the mean action of the entire fluid on the particle. 
(For a review, see, e.g., Ref. \cite{review}). 
In this case, the macroscopic wavefunction
for the condensate, $\Psi$, is determined 
by the Gross--Pitaevskii equation \cite{gp}
\begin{equation}
\label{nls}
i\hbar \frac{\partial \Psi}{\partial t} = - \frac{\hbar^2}{2m} \nabla^2 
\Psi + V({\rm \bf r}, t) \Psi +g | \Psi|^2\Psi  , 
\end{equation}
where $m$ is the mass of the atoms in the condensate, 
$V({\rm \bf r}, t)$ represents the trapping potential, and $g$ 
parametrizes the strength of the atomic interactions. 

Eq. (\ref{nls}) is 
formally equivalent to a three--dimensional,
non--linear Schr\"odinger equation and,  
in general, it is non--integrable. 
However, in two dimensions it is possible to determine 
the dynamics of the wavefunction by employing the 
`moment method' \cite{moment,moment1}. In this approach, integral 
relations are constructed directly from the wavefunction 
without solving the Schr\"odinger equation explicitly
and the evolution of these physical quantities 
then parametrizes the dynamics of the wavepacket. 

In view of this, we consider a two--dimensional, cylindrically symmetric 
Bose--Einstein condensate 
in a parabolic trapping potential 
$V(r ,t ) = m \omega^2r^2/2$ with a time--dependent 
frequency, $\omega = \omega (t)$. 
Lower--dimensional Bose gases are currently of interest both 
from the theoretical \cite{scaling,scaling1,moment1,2dtheo} 
and observational 
perspectives and  
two--dimensional condensates have recently been observed 
experimentally \cite{2dexpt}. 
A Bose--Einstein condensate becomes effectively two--dimensional 
when the excitations in a given direction are 
frozen and this occurs when the confining frequency in that direction 
is significantly greater than the energy for the thermal fluctuation.

We now briefly summarize the moment method \cite{moment,moment1a,moment1}. 
For the case of a constant interaction, $g$, 
the four integral parameters \cite{moment,moment1}
\begin{eqnarray}
\label{moment1}
I_1 (t) = \int d^2x | \Psi |^2  \\
\label{moment2}
I_2 (t) = \int d^2x r^2 | \Psi |^2 \\
\label{moment3}
I_3 (t) =i \int d^2x r \left[ \Psi \frac{\partial \Psi^*}{\partial r} 
- \Psi^* \frac{\partial \Psi}{\partial r} \right] \\
\label{moment4}
I_4 (t) = \frac{1}{2} \int d^2 x \left[ | \nabla \Psi |^2 + g 
| \Psi |^4 \right]
\end{eqnarray}
are defined, where it is assumed that $\Psi = \Psi (r ,t)$ 
and we have specified $\hbar =m=1$ without loss of generality. 
Eqs. (\ref{moment1})--(\ref{moment4}) admit physical interpretations in 
terms of the norm, width, radial momentum and 
energy of the wavepacket, respectively \cite{moment1}. 

By differentiating Eqs. (\ref{moment1})--(\ref{moment4}) in turn and 
substituting in Eq. (\ref{nls}), it can be shown that these integral 
parameters 
satisfy the set of coupled, first--order, ODEs \cite{moment1}: 
\begin{eqnarray}
\label{odemoment1}
\frac{dI_1}{dt} =0 \\
\label{odemoment2}
\frac{dI_2}{dt} = I_3 \\
\label{odemoment3}
\frac{dI_3}{dt} =-2\omega^2 (t) I_2 +4I_4 \\
\label{odemoment4}
\frac{dI_4}{dt} =- \frac{1}{2} \omega^2 (t) I_3  .
\end{eqnarray}
It may now be verified by direct differentiation 
that the 
quantity
\begin{equation}
\label{defQ}
Q \equiv 2I_4I_2 -\frac{I_3^2}{4} 
\end{equation}
is a 
constant and defining $X \equiv I_2^{1/2}$ then implies that the system 
(\ref{odemoment2})--(\ref{defQ}) is solved by solving 
the non--linear differential equation 
\begin{equation}
\label{pinney}
\frac{d^2 X}{dt^2} +\omega^2 (t) X= \frac{Q}{X^3}  .
\end{equation}

Eq. (\ref{pinney}) corresponds to the one--dimensional 
Ermakov system and is known as the Pinney (or Ermakov--Pinney) 
equation \cite{em,pinney}. 
The general solution of Eq. (\ref{pinney})
is given by \cite{pinney}
\begin{equation}
\label{pinneygensol}
X= \left[ AX_1^2+BX^2_2 +2CX_1 X_2 \right]^{1/2}  ,
\end{equation}
where $X_{1,2} (t)$ are linearly independent solutions to the 
equation
\begin{equation}
\label{linear}
\frac{d^2 X_i}{dt^2}  +\omega^2(t) X_i =0  ,
\end{equation}
the constants $\{ A,B,C \}$ satisfy the constraint 
\begin{equation}
AB-C^2 =\frac{Q}{W^2}
\end{equation}
and  
\begin{equation}
W \equiv X_1 \frac{dX_2}{dt} - X_2 \frac{dX_1}{dt}
\end{equation}
is 
the Wronskian. It follows, therefore, that the evolution of the wavepacket 
for a given trapping frequency, $\omega^2(t)$, 
is fully 
determined by solving the {\em linear} oscillator equation (\ref{linear}). 

We now proceed to consider 
cosmological dynamics. 
Our aim is to 
identify a class of models where
the expansion (or contraction) of the universe 
can be represented as an Ermakov system. 
Recently, it was shown that 
the Einstein field equations for a spatially flat, 
isotropic and homogeneous 
Friedmann--Robertson--Walker (FRW) universe 
can be reduced to an Ermakov--Pinney 
equation when the matter is a mixture of (non--interacting)
perfect fluids \cite{hl}. However, in this case, 
the sign of the non--linear term 
in the Ermakov--Pinney equation is opposite to 
that of Eq. (\ref{pinney}). This difference in sign arises 
because the energy densities of the matter fields  
are assumed to be positive--definite. 

In view of this, we consider the effects of 
introducing spatial curvature into the analysis. 
Einstein's field equations for the spatially curved FRW cosmology with  
perfect fluid matter are given by
\begin{eqnarray}  
\label{friedmann}
H^2 = \frac{1}{a^2} \left( \frac{da}{d\tau} \right)^2 = 
\frac{2}{3} \rho -\frac{k}{a^2} \\
\label{fluid}
\frac{d\rho}{d \tau} +3H (\rho +p ) =0  ,
\end{eqnarray}
where $H(\tau )$ is the Hubble parameter, $a (\tau )$ is the scale factor
of the universe,
$\tau$ represents cosmic (proper) time, $k$ is the curvature 
constant and units are chosen such that Newton's constant 
is given by $4\pi G=1$. The pressure and energy 
density of the matter are given by $p (\tau )$ and $\rho (\tau )$, 
respectively. 
In principle, 
Eqs. (\ref{friedmann}) and (\ref{fluid}) are solved once 
the equation of state, $p=p(\rho )$, has been specified. 
Eq. (\ref{fluid}) follows as a direct consequence of 
energy--momentum conservation. 

Eqs. (\ref{friedmann}) and (\ref{fluid}) 
may be expressed as an Ermakov system by 
interpreting the matter source as a 
self--interacting scalar field, $\phi = \phi (\tau )$, that is minimally 
coupled to Einstein gravity. Formally, this is equivalent  
to expressing the energy density and pressure as 
\begin{equation}
\label{scalar}
\rho = \frac{1}{2} \left( \frac{d\phi}{d\tau } \right)^2 +U(\phi ) , \qquad 
p= \frac{1}{2} \left( \frac{d \phi}{d \tau } \right)^2 -U(\phi )  ,
\end{equation}
where $U(\phi )$ represents the potential energy associated 
with the self--interactions of the scalar field. 
Differentiating Eq. (\ref{friedmann}) 
with respect to cosmic time and substituting 
Eqs. (\ref{fluid}) and (\ref{scalar}) then implies that 
\begin{equation}
\label{acc}
\frac{1}{a} \frac{d^2 a}{d \tau^2} - \frac{1}{a^2} \left( 
\frac{da}{d \tau } \right)^2 = -  \left( 
\frac{d \phi}{d\tau} \right)^2 + \frac{k}{a^2}
\end{equation}
and defining a new independent variable 
\begin{equation}
\label{timerel}
\frac{d}{d\tau} = a \frac{d}{dt}
\end{equation}
transforms Eq. (\ref{acc}) 
into the Pinney equation: 
\begin{equation}
\label{cosmicpinney}
\frac{d^2a}{dt^2} + \left( \frac{d\phi}{dt} \right)^2 a 
= \frac{k}{a^3}   .
\end{equation}

A direct comparison between Eqs. (\ref{pinney}) and (\ref{cosmicpinney})
indicates that the dynamics of 
a positively curved $(k>0)$ 
FRW cosmology can be modeled in terms of a harmonically trapped 
Bose--Einstein condensate when cosmic time, $\tau$, is related to 
`laboratory' time, $t$, through Eq. (\ref{timerel}). 
The width of the wavepacket plays the same role 
as that of 
the cosmological scale factor:
\begin{equation}
\label{I2}
I_2 \longleftrightarrow a^2
\end{equation}
and it follows from Eqs. (\ref{odemoment2}) and (\ref{timerel}) 
that the radial momentum of the wavepacket is associated with the 
expansion velocity of the universe: 
\begin{equation}
I_3 \longleftrightarrow 2\frac{da}{d\tau}  .
\end{equation}
This implies that the Hubble parameter may be identified with 
the combination
\begin{equation}
\label{I3}
H^2 \longleftrightarrow \frac{1}{4} \frac{I_3^2}{I_2}
\end{equation}
and consequently  the comoving Hubble radius is determined 
by the inverse of the wavepacket momentum: 
\begin{equation}
\label{comove}
\frac{1}{aH} \longleftrightarrow \frac{2}{I_3}  .
\end{equation}
Furthermore, 
substituting Eqs. (\ref{I2}) and (\ref{I3}) into Eq. (\ref{defQ}) 
and comparing the result with the Friedmann equation (\ref{friedmann}) 
implies that the energy of the condensate wavepacket corresponds to the 
energy density of cosmic matter
\begin{equation}
\label{I4}
I_4 \longleftrightarrow \frac{\rho}{3} ,
\end{equation}
where we identify $Q=k$. Finally, the 
kinetic energy of the scalar field is related to the 
trapping frequency of the condensate such that 
\begin{equation}
\label{kinetic}
\left( \frac{d\phi}{d\tau} \right)^2 \longleftrightarrow I_2 \omega^2
\end{equation}
and this implies that the pressure of the field 
is determined by 
\begin{equation}
p \longleftrightarrow I_2 \omega^2 -3I_4  .
\end{equation}

\begin{table}
\begin{center}
\begin{tabular}{||c|c||}
\hline \hline
Cosmological Parameter & Wavepacket Parameter  \\
\hline 
$a$                   & $I_2^{1/2}$   \\
$H^2$                 & $I_3^2/(4I_2)$ \\
$\rho$    & $3I_4$ \\
$p$  &  $I_2\omega^2 -3I_4$ \\
\hline \hline
\end{tabular}
\end{center}
\footnotesize{Table 1: 
The dynamical correspondence between a positively 
curved FRW cosmology with a perfect fluid 
matter source and a harmonically trapped, quasi--two--dimensional 
Bose--Einstein condensate. The width, radial momentum and 
energy of the wavepacket are  
analogous to the scale factor of the universe, the Hubble parameter and 
energy density of cosmic matter, respectively. 
The cosmic dynamics is determined once the relationship between the 
pressure and energy density has been specified. This is 
equivalent in the condensed matter context 
to choosing the time dependence of the trapping potential.}
\label{table1}
\end{table}

Thus, the width, radial momentum and energy of the 
condensate wavepacket can be identified with 
a corresponding cosmological parameter. This 
correspondence is summarized in Table 1. 
Eq. (\ref{defQ}) may now be viewed as a 
novel way of expressing the Friedmann equation in 
terms of properties of the wavefunction. Moreover, by employing the 
correspondences in Table 1, we deduce that 
Eq. (\ref{odemoment4}) is equivalent to the 
conservation equation (\ref{fluid}).

It is of interest to consider specific classes of cosmological 
models and their condensate analogues. 
For example, inflationary solutions are 
of particular importance to early universe cosmology. 
The condition for inflationary expansion is that the 
scale factor of the universe should accelerate (with respect to cosmic time) 
and it is straightforward to 
verify that this condition is satisfied whenever the 
radial momentum of the wavepacket grows: 
\begin{equation}
\label{inflation}
\frac{d^2a}{d \tau^2} >0 \qquad \Longleftrightarrow \qquad 
\frac{dI_3}{dt} >0 .
\end{equation}
Another important class of models is characterized by the 
barotropic equation of state, 
$p=(\gamma -1) \rho$, where $0 \le \gamma \le 2$ is a constant. 
In this case, 
the correspondence in Table 1 implies that 
the trapping potential satisfies $\omega^2=3\gamma I_4/I_2$
and substituting 
this condition into Eq. (\ref{odemoment4}) and integrating 
implies that 
\begin{equation}
\label{omegaI2}
I_4=\frac{A}{I_2^{3\gamma /2}} , \qquad \omega^2 = 
\frac{3\gamma A}{I_2^{(2+3\gamma )/2}} ,
\end{equation}
where $A$ is an arbitrary integration constant. 
A further 
substitution of Eqs. (\ref{odemoment2}) and 
(\ref{omegaI2}) into Eq. (\ref{defQ}) then 
yields a first--order differential equation for the width of the wavepacket:
\begin{equation}
\label{odeI2}
\frac{dI_2}{dt} = 2\sqrt{2AI_2^{(2-3\gamma )/2} -1} ,
\end{equation}
where we have specified $Q=1$ without loss of generality. 
Eq. (\ref{odeI2}) may be integrated to yield 
the solution 
\begin{eqnarray}
\label{2F1}
 {_{2}}F_1\left[ 
\frac{1}{2} \left(  \frac{3\gamma +2}{3\gamma -2} \right) , \frac{1}{2} ; 
\frac{1}{2} \left( \frac{9 \gamma -2}{3\gamma -2} \right) ; \frac{1}{2A} 
I_2^{(3\gamma -2)/2} \right] I_2^{(3\gamma+2)/4} \nonumber \\
= \sqrt{\frac{A}{2}} \left( 3\gamma +2 \right) 
\Delta t ,
\end{eqnarray}
where ${_{2}}F_1$ is the hypergeometric function 
\cite{as}, $\Delta t\equiv t-t_0$ and $t_0$ represents the 
second integration 
constant. 

Eqs. (\ref{omegaI2}) and (\ref{2F1})  
together determine the time dependence of 
the condensate trapping potential that mimics a cosmology with  
a given equation of state, $\gamma$. 
The limiting case of $\gamma =0$ represents a universe 
dominated by a cosmological constant and corresponds to a vanishing 
trapping potential. On the other hand, a universe dominated by 
relativistic matter has an equation of state given by 
$\gamma =4/3$ and in this case
Eq. (\ref{2F1}) simplifies 
to 
\begin{eqnarray}
\label{radiation}
{\sin}^{-1} \sqrt{x} -\sqrt{x-x^2} = \frac{1}{A} \Delta t 
\nonumber \\
x \equiv \frac{(4A)^{1/3}}{2A} \frac{1}{\omega^{2/3}}  .
\end{eqnarray}
Eq. (\ref{2F1}) also simplifies for a universe dominated by pressureless 
matter $(\gamma =1)$: 
\begin{eqnarray}
\label{dust}
3{\sin}^{-1} \sqrt{y} -(3+2y ) \sqrt{y-y^2}  = \frac{1}{A^2} \Delta t
\nonumber \\
y \equiv \frac{(3A)^{1/5}}{2A} \frac{1}{\omega^{2/5}} .
\end{eqnarray} 
Although 
Eqs. (\ref{radiation}) and (\ref{dust}) can not 
be inverted analytically, the explicit time dependence 
of the trapping potential can be expressed in a closed form for 
a `stiff' perfect fluid $(\gamma =2)$. We find in this case that 
\begin{eqnarray}
\label{gamma2}
I_2 = \left[ 2A-4 (\Delta t)^2 \right]^{1/2} 
\nonumber \\
\omega^2 = \frac{3A}{2 \left[ A-2(\Delta t)^2 \right]^2}  .
\end{eqnarray}

To summarize thus far,
it has been shown that  
positively curved, 
perfect fluid FRW cosmologies can be modeled dynamically 
in terms of 
quasi--two--dimensional Bose--Einstein condensates, 
where there exists a one--to--one correspondence between
the type of matter in the universe and the functional form of 
the time--dependent trapping potential of the condensate. The
physical properties of the 
wavefunction can be identified with the 
fundamental cosmological parameters. 

It is of interest to consider extensions of the 
correspondence to three--dimensional condensates. 
The moment method as outlined above can not 
be applied in three dimensions unless certain 
approximations are made \cite{moment1a}. However, 
an alternative approach is to employ the 
well known scaling properties that are exhibited by the wavefunction
under the evolution of a time--dependent 
confining potential \cite{scaling,scaling1,scaling2}.
In this approach, the 
{\em ansatz} 
\begin{equation}
\label{ansatz}
\Psi = \frac{1}{b^{3/2}} \chi \exp \left[ \frac{i}{2} \frac{b_t}{b} 
r^2 \right]
\end{equation}
is made, where $b=b(t)$, $\chi =\chi (t, r )$ 
and a subscript denotes differentiation with respect to 
laboratory time, $t$ \cite{scaling}.
Substituting Eq. (\ref{ansatz}) into the non--linear Schr\"odinger 
equation (\ref{nls}) and
defining new variables 
\begin{equation}
\label{sigmadef}
\sigma \equiv \frac{r}{b} \qquad 
\theta \equiv \int \frac{dt}{b^2}
\end{equation}
then implies that Eq. (\ref{nls}) reduces 
to 
\begin{eqnarray}
\label{reduce}
\frac{i}{b^2} \frac{\partial \chi}{\partial \theta} 
+\frac{1}{2b^2} \nabla^2_{\sigma} \chi -\frac{g}{b^3} 
| \chi |^2 \chi
-\frac{1}{2} \frac{\omega_0^2\sigma^2\chi}{b^{2}} 
\nonumber \\ 
= 
\frac{1}{2} r^2 \chi \left[ \frac{1}{b} \frac{d^2b}{dt^2} +\omega^2(t) -
\frac{\omega_0^2}{b^{4}} \right]  ,
\end{eqnarray}
where $\omega_0$ is a constant and the last terms on both sides 
of Eq. (\ref{reduce}) have been introduced by hand. 

If the interaction strength, $g$, is time--independent, 
it is not possible in general 
to separate the left and right hand sides of Eq. 
(\ref{reduce}). On the other hand, the scattering length may be varied 
by employing a magnetic field induced
Feshbach resonance \cite{fesh}.
If the interaction is tuned such that 
$g(t) \propto b(t)$, all terms on the left--hand side of 
Eq. (\ref{reduce}) scale as $b^{-2}$ and, 
in this case, Eq. (\ref{reduce}) 
separates when $b(t)$ satisfies 
the Ermakov--Pinney equation 
\begin{equation}
\label{quarticpinney}
\frac{d^2 b}{dt^2} +\omega^2 (t) b =\frac{\omega_0^2}{b^3}  .
\end{equation}
It then follows that 
the rescaled wavefunction, $\chi$, formally satisfies the 
same non--linear Schr\"odinger equation as the original 
wavefunction but with a constant trapping frequency and interaction strength. 
Comparing Eqs. (\ref{cosmicpinney}) and (\ref{quarticpinney}) 
implies that the scaling parameter, $b$,  may be identified with the scale 
factor of the universe, where laboratory time is again related to cosmic 
time through Eq. (\ref{timerel}). 

We conclude by highlighting a number of questions that would 
need to be addressed in order 
to realise 
the above mathematical correspondences in a laboratory 
environment. In general, 
a sonic horizon may 
form in an expanding Bose--Einstein condensate, 
where the speed of the fluid exceeds the sound speed. 
Such an horizon forms due to an 
effective velocity field 
that increases with distance from the centre 
of the condensate \cite{blv}. This 
could be problematic from a dynamical 
point of view since it may introduce physical effects 
into the system that may not be intrinsic to the 
cosmological model under consideration. In particular, 
ambiguities may arise when interpreting the analogue of cosmological 
particle production  in a time--dependent gravitational field. 

A key assumption that was made in
establishing the correspondence between the condensate and 
cosmological systems was that the dynamics of the 
condensate wavefunction can be described in terms of 
the Gross--Pitaevskii equation 
(\ref{nls}) at each moment of time, 
i.e., that the 
configuration reacts instantaneously to changes in the trapping potential and 
scattering length of the atomic interactions. If this assumption is 
to remain valid,
the majority of the atoms must remain in the condensate 
state (mean--field approximation) and 
the particle density and scattering length must be 
sufficiently small (dilute gas approximation).  
Moreover,  
to avoid complications with delays in the system, 
the timescales over which external parameters such as the trapping potential 
change should 
exceed the timescale for a pair of atoms to fully
interact \cite{blv,kb}. 
Overall, these requirements imply that the 
trapping potential should change at a sufficiently 
slow rate. In this case, 
the correspondence (\ref{kinetic}) implies that 
only those cosmologies where the kinetic 
energy of the scalar field is sufficiently small could be 
simulated\footnote{It should be noted, however, that in an inflationary 
context, the scalar field is necessarily rolling very slowly down 
its interaction potential.}. 

A further important scale in the 
condensate is the `healing' time, $t_{\rm heal}$. 
The healing time is the characteristic timescale for variations 
in the wavefunction and  
plays an analogous role to that of the Planck time 
in cosmology. Comparison of cosmic and laboratory time through the 
correspondence (\ref{timerel}) implies that 
$a d\tau = dt > t_{\rm heal}$ and, 
for a cosmic timescale of the order of 
the Hubble time $d \tau = H^{-1}$ (representing  
the size of the cosmic horizon), this implies that  
$a>t_{\rm heal} H$. Such considerations suggest that 
the correspondence we have developed could only be realised in practice 
for cosmological models satisfying a lower bound on the scale factor. 

Nevertheless, one advantage of
establishing correspondences between cosmology and 
condensed matter physics through Ermakov systems 
is that insight into the hidden symmetries 
of the two systems may be uncovered. 
For example, in the absence of 
a trapping potential, the action for a two--dimensional 
Bose--Einstein condensate described by the non--linear 
Schr\"odinger equation (\ref{nls}) is symmetric under a 
global ${\rm SL}(2,R)$ 
reparametrization of the time parameter:  $\tilde{t} = 
(\alpha t+ \beta )/(\gamma t + \delta )$, where $\{ \alpha , \beta 
, \gamma , \delta \}$ are arbitrary constants satisfying the constraint
$\alpha \delta -\beta \gamma =1$ \cite{gh,gh1}. In general, 
this symmetry 
is broken by the inclusion of a time--dependent 
trapping potential. However, for the specific case where the potential 
varies as 
$\omega (t) \propto t^{-2}$, there exists a discrete duality 
symmetry with $\alpha =\delta =0$ and $\beta = -1/\gamma $ \cite{gh1}. 
This implies that
$\tilde{t} = -\beta^2 /t$ and such a transformation 
maps an expanding solution  
onto a contracting one and vice--versa. 

The associated cosmological 
model for this trapping potential 
may be deduced in the limit where the spatial curvature is 
dynamically negligible, as is the case during inflation. 
It follows  
from the 
correspondence summarized in Table 1 that for this 
trapping potential the scale factor has a power law dependence, 
$a (\tau) \propto \tau^n$ for 
some constant $n$, and the scalar field varies as 
$\phi (\tau) \propto \ln \tau$. This 
solution plays a central role in inflationary cosmology 
because it is one of the 
few models where the density and gravitational wave perturbation
spectra can be calculated exactly without 
invoking the slow--roll approximation \cite{lmsl} (for a review, see, e.g., 
Ref. \cite{llkcba}). In effect, the solution is a scaling 
solution since the kinetic energy of the scalar 
field and the Hubble parameter decrease at the same rate. 
In view of its significance, 
it is interesting that this cosmological model is selected 
directly from symmetries exhibited by a class of 
Bose--Einstein condensates. It would be interesting to 
investigate whether an associated symmetry exists 
in the cosmological context as this
would provide further insight 
into the density and gravitational wave perturbations 
generated during inflation.  

Finally, an ${\rm SL} (2,R)$ symmetry also arises in 
superstring--inspired 
inflationary cosmologies such as the pre--big bang scenario \cite{pbb}, where 
a duality symmetry relates contracting and expanding 
solutions. (For reviews, see, e.g., Ref. \cite{lwc}). In these  
models, the dilaton and axion fields in the Neveu--Schwarz/Neveu--Schwarz
sector of the string effective action parametrize 
the ${\rm SL}(2,R)/{\rm U}(1)$ coset and the existence 
of this symmetry implies that the density perturbations arising
from fluctuations in these fields can be calculated 
exactly \cite{cew}. Furthermore, inflation in pre--big bang 
cosmology can be re--interpreted as the contracting phase of a 
universe dominated by a stiff perfect fluid \cite{gv1}
and this corresponds in the context of the present work to the 
condensate given in Eq. 
(\ref{gamma2}).

\vspace{.7in}
{\bf Acknowledgments} 
\vspace{.3in}

JEL is supported by the Royal Society. We thank C. Barcel\'o, 
P. K. Ghosh and U. R. Fischer for 
helpful comments and communications.

\end{document}